\documentclass[aps,prb,superscriptaddress,twocolumn,showpacs,10pt]{revtex4-1}
\usepackage [utf8]{inputenc} 
\usepackage{graphicx, amsmath, nicefrac, color, ulem}

\begin{document}

\title{Conductance channels of a platform molecule on Au(111) probed with shot noise}

\author{Michael~Mohr} \affiliation{Institut f\"ur Experimentelle und Angewandte Physik, Christian-Albrechts-Universit\"at zu Kiel, D-24098 Kiel, Germany}
\author{Torben~Jasper-Toennies} \affiliation{Institut f\"ur Experimentelle und Angewandte Physik, Christian-Albrechts-Universit\"at zu Kiel, D-24098 Kiel, Germany}
\author{Alexander Weismann} \affiliation{Institut f\"ur Experimentelle und Angewandte Physik, Christian-Albrechts-Universit\"at zu Kiel, D-24098 Kiel, Germany}
\author{Thomas~Frederiksen} \affiliation{Donostia International Physics Center (DIPC), Paseo Manuel de Lardizabal 4, Donostia-San Sebasti\'an, Spain} \affiliation{IKERBASQUE, Basque Foundation for Science, Bilbao, Spain}
\author{Aran~Garcia-Lekue} \affiliation{Donostia International Physics Center (DIPC), Paseo Manuel de Lardizabal 4, Donostia-San Sebasti\'an, Spain} \affiliation{IKERBASQUE, Basque Foundation for Science, Bilbao, Spain}
\author{Sandra~Ulrich} \affiliation{Otto-Diels-Institut f\"ur Organische Chemie, Christian-Albrechts-Universit\"at zu Kiel, 24098 Kiel, Germany}
\author{Rainer~Herges} \affiliation{Otto-Diels-Institut f\"ur Organische Chemie, Christian-Albrechts-Universit\"at zu Kiel, 24098 Kiel, Germany}
\author{Richard~Berndt} \affiliation{Institut f\"ur Experimentelle und Angewandte Physik, Christian-Albrechts-Universit\"at zu Kiel, D-24098 Kiel, Germany}

\begin{abstract}
The shot noise of the current $I$ through junctions to single trioxatriangulenium cations (TOTA$^+$) on Au(111) is measured with a low temperature scanning tunneling microscope using Au tips.
The noise is significantly reduced compared to the Poisson noise power of $2eI$ and varies linearly with the junction conductance.
The data are consistent with electron transmission through a single spin-degenerate transport channel and show that TOTA$^+$ in a Au contact does not acquire an unpaired electron.
Ab initio calculations reproduce the observations and show that the current involves the lowest unoccupied orbital of the molecule and tip states close to the Fermi level.
\end{abstract}


\date{\today}

\maketitle

\section{Introduction}

The shot noise of electrical current may be used to extract information on the transport process that is not available from the conductance alone as demonstrated on mesoscopic structures \cite{blanter}.
Some data are also available from atomic junctions \cite{VanDenBrom_PRL, RBG_Kumar_PRL, Schoenenberger, Scheer_Ir, kemiktarak_radio-frequency_2007, chen_enhanced_2014, Urbina_Al, Vardimon, vardimon2015indication, vardimon2016orbital, MKu, nonoise, abu, abuss}.
As to molecules, a few systems have so far been experimentally investigated with mechanically controlled breakjunctions.
From D$_2$ \cite{RBG_Djukic_single_molecule}, benzene \cite{Ruitenbeek_Benzene}, and H$_2$O \cite{talwasser}, an inelastic contribution to the conductance was reported.
For D$_2$ \cite{chain}, vanadocene \cite{Pal_2018} and C$_6$H$_4$S$_2$ \cite{karimi2016shot}, insight into the conducting quantum states was obtained.
High-frequency (sub-PHz) noise from C$_{60}$ junctions was interpreted in terms of transient charging of the molecule \cite{c60}.

\begin{figure}[hbt]	
		\includegraphics[width=1\columnwidth]{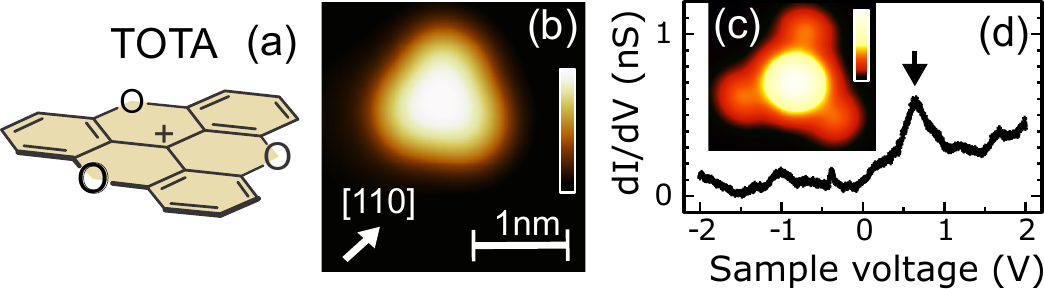}
		\caption{
		(a) Model of the calculated gas-phase structure of trioxatriangulenium (TOTA).
		(b) Constant-current STM topograph of a single TOTA molecule, which appears 
		    2.2\,\AA\ high at the sample voltage $V = 129$\,mV and the tunneling current $I = 28$\,pA\@.
		    A $\langle110\rangle$ direction of the Au substrate is indicated by an arrow.
		(c) Constant-height $dI/dV$  map of the LUMO of a single TOTA molecule recorded at $V=0.6$\,V\@.
    (d) Spectrum of the differential conductance $(dI/dV)$ of TOTA recorded above the center of the molecule.
        The feedback loop was opened at 500\,pA and 2\,V\@.}
\label{stm}
\end{figure}

Here, we discuss experimental and theoretical results for trioxatriangulenium (TOTA$^+$, C$_{19}$H$_9$O$_3$, Fig.~\ref{stm}a), an aromatic cation, adsorbed to Au(111).
In solution, TOTA is planar and evolves to a bowl shape when a functional unit is binding to its central carbon atom \cite{mands, Baisch_2009}.
TOTA is used as a platform for molecular wires to enable reproducible conductance measurements in a low-temperature scanning tunneling microscope (STM) \cite{tjtwire, tjtcond}.
It is conceivable that the molecule becomes neutral on a metal surface and therefore may contain an unpaired electron.
When contacted with a Au atom at a STM tip, TOTA on the Au(111) surface exhibits a high conductance of $\approx 0.4$\,G$_0$, where G$_0= 2e^2/h$.
With the help of transport calculations we assign this conductance, which is fairly large for an organic molecule, to the electron transport through a single transmission channel that is derived from the lowest unoccupied molecular orbital (LUMO) of the molecule on the surface.
In contact, where a Au-C bond is formed, the LUMO state strongly hybridizes with tip states.
Moreover, we present shot noise data that exclude significant spin-polarisation of the transport through this resonance.
The calculated shot noise matches the experimental data, which further excludes the possibility that TOTA is a radical when contacted with Au electrodes.

\section{Experimental details}

All experiments were carried out in ultra-high vacuum (UHV).
We used a custom made low-temperature STM operated at 4.5~K\@.
A Au(111) substrate was cleaned by cycles of Ar sputtering and annealing.
Sub-monolayer coverages of TOTA were obtained by sublimation onto the sample at ambient temperature from a Knudsen cell with the salt TOTA$^+$BF$_4^-$.
We also prepared TOTA molecules by dissociating adsorbed methyl-TOTA molecules with the help of current pulses.
The results reported below were identical for both methods of preparation.
Further details on the preparation may be found in Ref.~\onlinecite{tjtstab}.
Low coverages were used to enable tip preparation on clean substrate areas.
Tips from electrochemically etched W wire were prepared in UHV by annealing and indentation into the Au crystal.
The tips are thus coated with gold as verified by depositing single Au atoms from the tip.
We made sure that the measurements on molecules did not change the tips.

We used the setup described in Ref.~\cite{abu} to measure the noise of the current.
Once a molecular contact was prepared, the conventional STM control electronics was disconnected and current was driven through the junction and two series resistors with a battery-powered circuit.
The voltage noise signal at the junction was amplified and cross-correlated to reduce the effect of amplifier noise.
This leads to a spectrum of the noise power density, which we evaluated in a frequency range from 50 to 200~kHz.
Pink noise ($1/f^\alpha$) and cable capacitances presently prevent useful measurements outside of this range.
We only used the data if the conductance did not exhibit significant changes throughout the measurements at typically 9 current values.
In addition to the noise, we also determined the DC voltage drop over the contact to calculate its conductance. 

\section{STM of TOTA on Au(111)}

In constant-current STM images (Fig.~\ref{stm}b), the molecule appears as a triangular protrusion, as expected owing to its threefold symmetry, with an apparent height of $\leq 2.2$~{\AA}.
TOTA adsorbs flat on metals and binds via its extended conjugated $\pi$ system \cite{tjtstab}.
A map of the differential conductance $dI/dV$, that was recorded at constant height of the tip above the Au substrate using a sample voltage $V=0.6$\,V is displayed in Fig.~\ref{stm}c.
Being related to the local density of states, this map reveals a resonance that is mainly located at the center of the molecule.
A $dI/dV$ spectrum recorded above the center of the molecule (Fig.~\ref{stm}d) shows that this resonance is fairly sharp and centered approximately 0.6\,eV above the Fermi level.

\section{Experimental contact conductance}

An electrical contact to a TOTA molecule may be reproducibly formed by imaging with the STM, centering the tip above the molecule, disabling the current feedback, and then reducing the tip-molecule distance.
For these measurements we used single molecules adsorbed to the elbows of the herring bone reconstruction because other single molecules tended to lead to instabilities of the junction.
Besides this stability aspect we observed no significant differences between single molecules.
In particular, the conductance data were essentially the same as long as stable contacts were formed.

\begin{figure}[hbt]	
		\includegraphics[width=0.75\columnwidth]{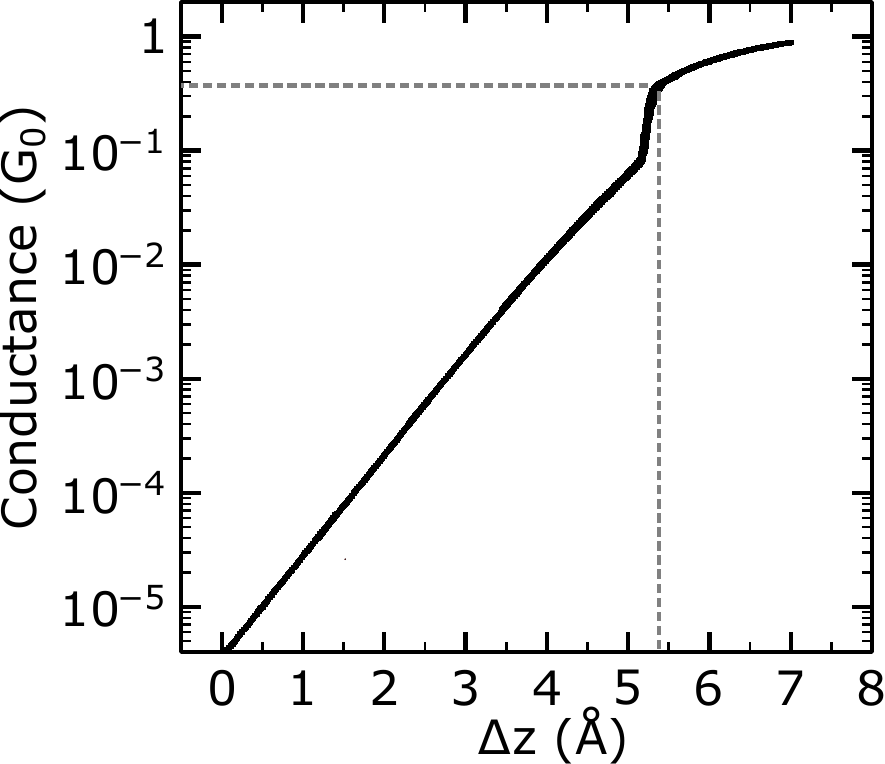}
		\caption{
		Measured conductance $G$ of a TOTA junction vs.\ vertical displacement $\Delta z$ of the STM tip.
		The tip was centered above the molecule and a sample voltage $V=0.129$~V was used.
    $\Delta z=0$ corresponds to the tip position at $I=40$~pA\@.
		Contact with the same tip to an fcc area of the bare substrate occurs at $\approx 8$~\AA\@.
		The data shown were measured during 7 approaches and retractions of the tip.
		Each approach or retraction consisted of overlapping pieces that extended over approximately 2 decades in conductance. 
		The reproducibility of the experimental result is indicated by the resulting line thickness.}
\label{gz}
\end{figure}

Figure~\ref{gz} shows data that cover the tunneling range (small $\Delta z$) to contact.
Here we used the rapid rise of the conductance at $\Delta z \approx 5.4$\,\AA\ as an indicator of contact formation.
Right after contact has been reached, the conductance is $\approx 0.4$\,G$_0$.
It keeps increasing beyond this point, albeit with a reduced slope.
It should be noted that the data shown in Fig.~\ref{gz} were actually recorded both during the approach and the retraction of the tip.
Moreover, the results of 7 repeated measurements are simultaneously presented.
This makes obvious the absence of hysteresis and the reproducibility of the measurements.

\section{Shot noise measurements}

To measure spectra of the current noise as a function of the current, $S(I)$, we used molecules adsorbed to the elbows of the herring bone reconstruction.
The spectra were treated as follows.
The noise is the sum of the excess noise $\Delta S$ and the thermal noise $S_\theta = 4 k_B T G$ ($T$ temperature, $k_B$ Boltzmann constant).
$\Delta S = F\left[ S_0 \, \coth\left({S_0}/{S_{\theta}}\right)-S_{\theta} \right]$ is related to the Fano factor $F$ and the temperature~\cite{abu}.
$S_0=2 e I$ is the Poisson noise.
Noise data $S(I)$ recorded at a number of different currents including $I=0$ were fitted treating $F$ and $T$ as adjustable parameters.

\begin{figure}[hbt]	
		\includegraphics[width=0.8\columnwidth]{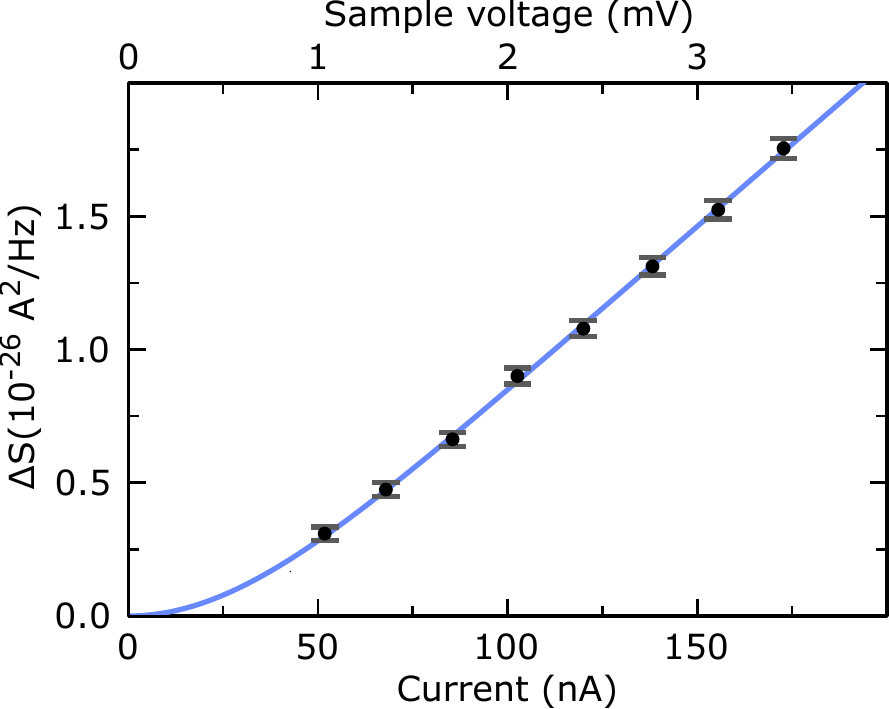}
		\caption{
		Noise power $\Delta S(I)$ measured on a TOTA molecule.
		Markers show experimental data with uncertainty margins. 
		The uncertainties were obtained from 400 measurements between 100 and 120~kHz.
		The line is the result of a fit $\Delta S = F\left[ S_0 \, \coth\left({S_0}/{S_{\theta}}\right)-S_{\theta} \right]$.
    The upper axis shows junction voltages.}
\label{exc}
\end{figure}

Figure \ref{exc} shows an example experimental data (dots) and the fit (line) obtained.
As expected at large $I$, $\Delta S$ evolves linearly with the current $I.$	
We note that the temperature determined from the thermal noise $S_\theta$ deviated less than $5~\%$ from the temperature of the STM measured with a diode.

\begin{figure}[hbt]	
		\includegraphics[width=0.9\columnwidth]{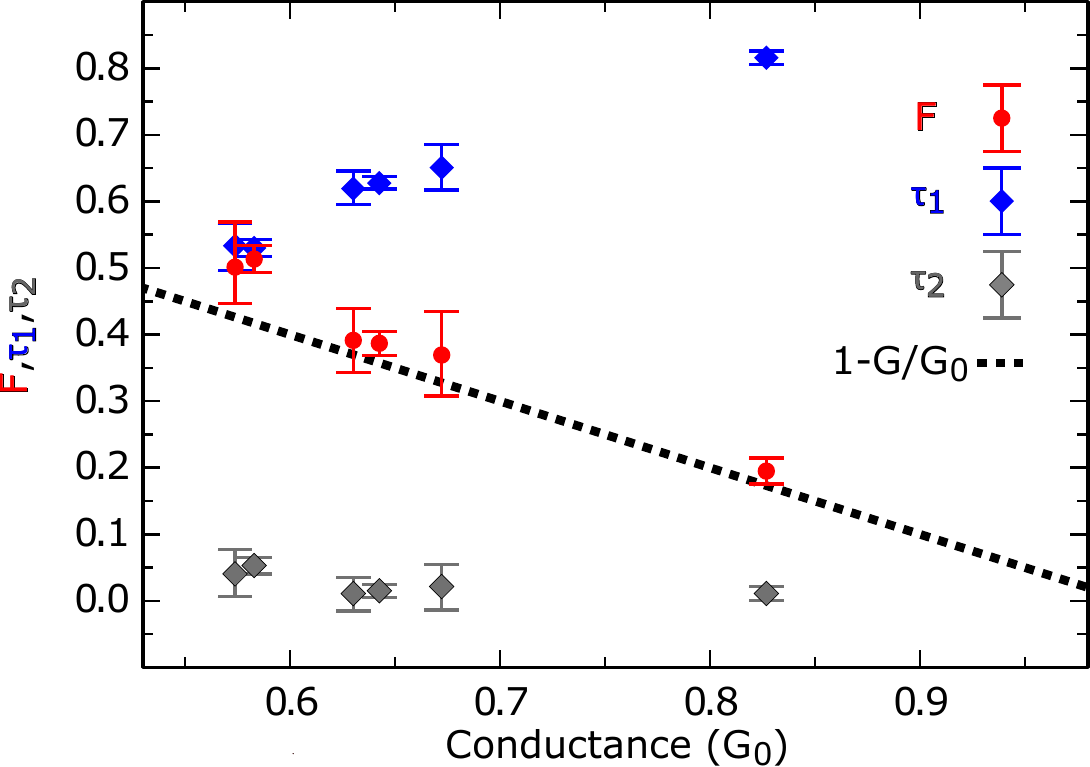}
		\caption{
		Experimental Fano factors $F$ (red cirles) and corresponding channel transmissions $\tau_1$ (blue diamonds) and $\tau_2$  (gray diamonds) vs.\ the conductances $G$ of trioxatriangulenium junctions.
		The data were measured from seven different contacts with possibly different atomic-scale structures of the tips.
		A dotted line shows the Fano factors expected from a model assuming a single spin degenerate transport channel ($1-G/\mathrm{G_0}$).
		Uncertainty margins are defined as 95\,\% confidence intervals. 
		The uncertainties of $G$  are negligible, \textit{i.\,e.}\ smaller than the marker sizes.}
\label{fanoxp}
\end{figure}

The Fano factors obtained for various junction conductances are shown in Fig.\,\ref{fanoxp} (red dots).
The data were measured in the contact range with $G$ varying between 0.4 and 0.8\,G$_0$, \textit{i.\,e.}\ at tip displacements somewhat beyond the point of contact formation.
For comparison, a dotted line shows the Fano factor expected in a model where a single, spin degenerate transport channel dominates the conductance.
In this case, $F=1-G/\mathrm{G}_0$ \cite{blanter}.
Except for some scatter, the measurements are close to this line.
As a simple model, we assume two spin-degenerate conductance channels with transmissions $\tau_1$ and $\tau_2$.
Using the relations $G/\mathrm{G}_0= \tau_1 + \tau_2$ and $F = \sum{\tau_i (1- \tau_i)}/\sum{\tau_i}$ \cite{blanter} we obtain the transmissions indicated by diamonds in Fig.\,\ref{fanoxp}.
The transmission of the second channel is at most a few percent.

\section{Theoretical methods}
	
The conductance properties of the TOTA junction were also studied with density functional theory (DFT) \cite{siesta2002} and nonequilibrium Green functions with \cite{Brandbyge2002,Papior2017} following the same computational approach as used previously \cite{tjtwire, tjtcond}, taking into account dispersion interactions \cite{Klimes2010}. 
The supercell (Fig.~\ref{theo}a) contained a single TOTA molecule adsorbed on a 10-layer Au(111) slab with $6 \times 6$ periodicity.
On the reverse side of the slab a 10-atom Au(111) tetrahedron was mounted to represent the STM tip apex.
The molecule, tip, and surface layer atom coordinates were relaxed until residual forces were smaller than 0.04\,eV/{\AA} (64\,pN). 
Projected density of states (PDOS) onto the the molecule and the tip and energy-resolved transmission $T(E)$ were computed as averages $\langle \cdots \rangle_k$ over the transverse Brillouin zone on a $21\times 21$ $k$-grid. 
The Fano factor was computed with periodic boundary conditions from the transmissions $\tau_i$ at the Fermi energy as $F=\langle \sum_i\tau_i(1-\tau_i)\rangle_k/\langle \sum_i \tau_i\rangle_k$ using Inelastica \cite{Frederiksen2007, Inelastica}.
As discussed in the Supplementary Information to Ref.~\cite{abu}, the $k$-point sampling, which is necessary because periodic boundary conditions are used in the calculations, leads to a slight deviation (reduction) of the Fano factor from the value expected for a single junction.

\section{Calculated conductances and Fano factors}

\begin{figure*}[hbt]	
	\includegraphics[width=0.9999\linewidth]{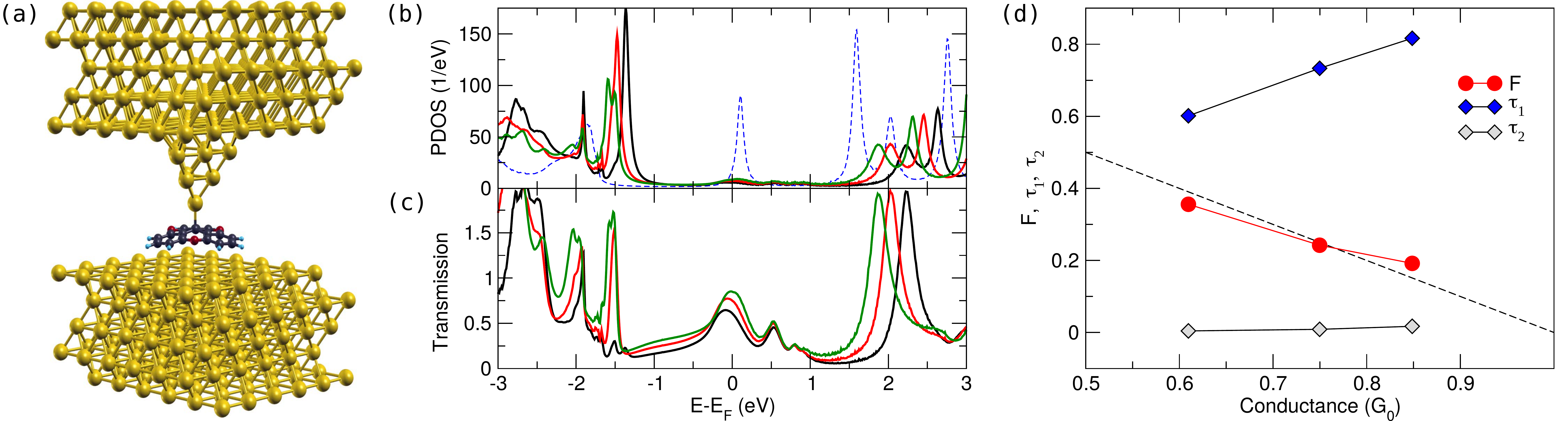}
	\caption{Transport simulations for a TOTA junction. 
	(a) Supercell containing one TOTA molecule on $6\times 6$ Au(111) surface. 
	Contact with a Au tip is through a bond to the central carbon atom.
	(b) Projected density of states onto the atoms of the tip and the molecule.
	Black, red, and green curves correspond to decreasing tip heights (steps of 0.5\,\AA) in the contact regime.
	The dashed blue curve corresponds to TOTA relaxed on the surface without interactions with the tip, revealing the LUMO resonance close to the Fermi level.
	(c) Electron transmission functions $T(E)$ and 
	(d) Fano factors along with the $k$-averaged channel transmissions $\tau_1$ and $\tau_2$ for the same three contact geometries.}
\label{theo}
\end{figure*}

Tunneling through the TOTA molecule is dominated by the LUMO resonance close to the Fermi level (dashed blue curve in Fig.~\ref{theo}b).
Its calculated energy is slightly lower than the peak energy of the experimental $dI/dV$ data (Fig.~\ref{stm}c) as expected in DFT calculations, which tend to underestimate HOMO--LUMO gaps \cite{Godby_1986}. 
In the contact regime, when the apex Au atom binds to the center carbon atom, this LUMO state strongly hybridizes with the tip states -- drastically changing the spectrum -- and electron transport proceeds through these hybridized tip-molecule states (Fig.~\ref{theo}c).

Around the Fermi energy we find that the TOTA junction is clearly dominated by a \textit{single} conductance channel.
Thus, the channel transmissions $\tau_1$ and $\tau_2$ can easily be calculated as averages over all $k$-points (blue and grey diamonds in Fig.~\ref{theo}d).
The computed Fano factors (red dots in Fig.~\ref{theo}d) closely follow the single-channel result.

\section{Discussion}

The match of the calculated transmissions and Fano factors with the experimental results is remarkably good.
The conductance of the Au-TOTA juctions may therefore safely be assigned to a single transport channel.
According to the calculations this eigenchannel exhibits $\sigma$ symmetry.
$\pi$-type channels contribute but a few percent of the conductance.
First, the Au atom at the tip apex with its predominant $s$-character induces a degree of orbital filtering.
Second, the LUMO state (dashed blue curve in Fig.~\ref{theo}b) around the central C atom of TOTA is dominated by the $p_z$ orbital.
Upon formation of the Au-C bond, the dominant channel remains of $\sigma$-type.
Consequently, the higher angular momentum conductance channels are suppressed.
Indeed, calculations show that an electron wave incoming from the substrate side in the second ($\pi$-type) eigenchannel at $E_F$ is already reflected at the substrate-molecule interface.
The conductance and noise data recorded in the contact range show that the electron transmission is spin-degenerate.
TOTA$^+$ has no unpaired electrons in the gas phase and the contact with two Au electrodes (tip and sample) does not appear to change this state of affairs.
In agreement with this interpretation, spectra of the differential conductance near the Fermi level recorded at contact (not shown) are featureless.
Possible indications of a localized spin, e.\,g.\ via the Kondo effect or spin excitations, are not observed from the molecular contacts.

\section{Summary}

Using trioxatriangulenium cations deposited on Au(111) we demonstrated that STM-based measurements of shot noise may be extended to contacts to largish molecules.
The observed shot noise and conductance data are reproduced by transport calculations and reflect electron transport through a single, spin degenerate channel.

\begin{acknowledgments}
We thank the Deutsche Forschungsgemeinschaft (SFB 677) and the Spanish MINECO (Grants No.\ MAT2016-78293-C6-4-R and FIS2017-83780-P) for financial support.
\end{acknowledgments}

%

\end{document}